\documentclass[12pt]{iopart}

\usepackage{iopams} 
\usepackage{verbatim}
\usepackage{graphicx}
\usepackage{color}
\begin{document}

\title{Ternary Stochastic Neuron -- Implemented with a Single Strained Magnetostrictive Nanomagnet}

\author{Rahnuma Rahman and Supriyo Bandyopadhyay}

\address{Department of Electrical and Computer Engineering, Virginia Commonwealth University, Richmond, VA 23284, USA}
\ead{sbandy@vcu.edu}
\vspace{10pt}

\begin{abstract}
Stochastic neurons are extremely efficient hardware for solving a large class of problems and usually come in two varieties -- ``binary'' where the neuronal state varies randomly between two values of $\pm$1 and ''analog'' where the neuronal state can randomly assume any value between -1 and +1.  Both have their uses in neuromorphic computing and both can be implemented with low- or zero-energy-barrier nanomagnets whose random magnetization orientations in the presence of thermal noise encode the binary or analog state variables. In between these two classes is $n$-ary stochastic neurons, mainly ternary stochastic neurons (TSN) whose state randomly assumes one of three values (-1, 0, +1), which have proved to be efficient in pattern classification tasks such as recognizing handwritten digits from the MNIST data set or patterns from the CIFAR-10 data set. Here, we show how to implement a TSN with a zero-energy-barrier (shape isotropic) magnetostrictive nanomagnet subjected to uniaxial strain.

\end{abstract}

\medskip


%
%
%
\maketitle
%
%

\section{Introduction: Ternary stochastic neurons (TSNs)}

Deep neural networks (DNN) are essential ingredients of artificial intelligence (AI), but deploying them at the edge or for embedded applications (wearable electronics, personal communicators, etc.) faces significant challenge because of the need for massive storage and computational power. Numerous compression techniques have been proposed to reduce the computational burden \cite{rastegari,esser,han,courbariaux,hubara,lin,iandola,howard,zhang,liu,wang1,wang2} and have performed admirably well. A different approach that departs from conventional compression methods is to use {\it ternary quantization} to reduce storage and computational cost \cite{li}.  Here every neuron has three states instead of the usual two in binary quantization. Hence, there is an obvious saving in area and energy cost which is beneficial for edge applications. Ternary quantization has been shown to perform very well in image classification and detection tasks involving MNIST, CIFAR-10 and {\bf Imagenet} \cite{li,zhu}. It also has very little accuracy degradation and can improve the accuracy of some models \cite{zhu}. When both the weights and activations assume ternary values, the DNNs can be reduced to sparse binary networks with significant space and energy savings \cite{deng}.

Ever since the dawn of digital electronics, binary devices with two states to encode digital information have ruled the roost. While they are easy to implement, they provide the least information content per device. As a result, even in the digital logic community, there is a strong push to switch to a higher radix/logic number by employing multi-state logic \cite{zahoor}.  If the same device can encode $n$ states ($n > 2$)instead of two, it obviously can process more data in the same area of a chip, resulting in improvements in {\it computational density}. When it comes to low power and energy consumption, low complexity,
low on-chip and off-chip interconnections, and high speed, the $n$-ary system ($n > 2$) is inevitably superior to the binary system \cite{zahoor}. This is true in all forms of information processing. 

\begin{figure}[!ht]
\centering
\vspace{-0.4cm}
\includegraphics[width=4.3in,angle=270]{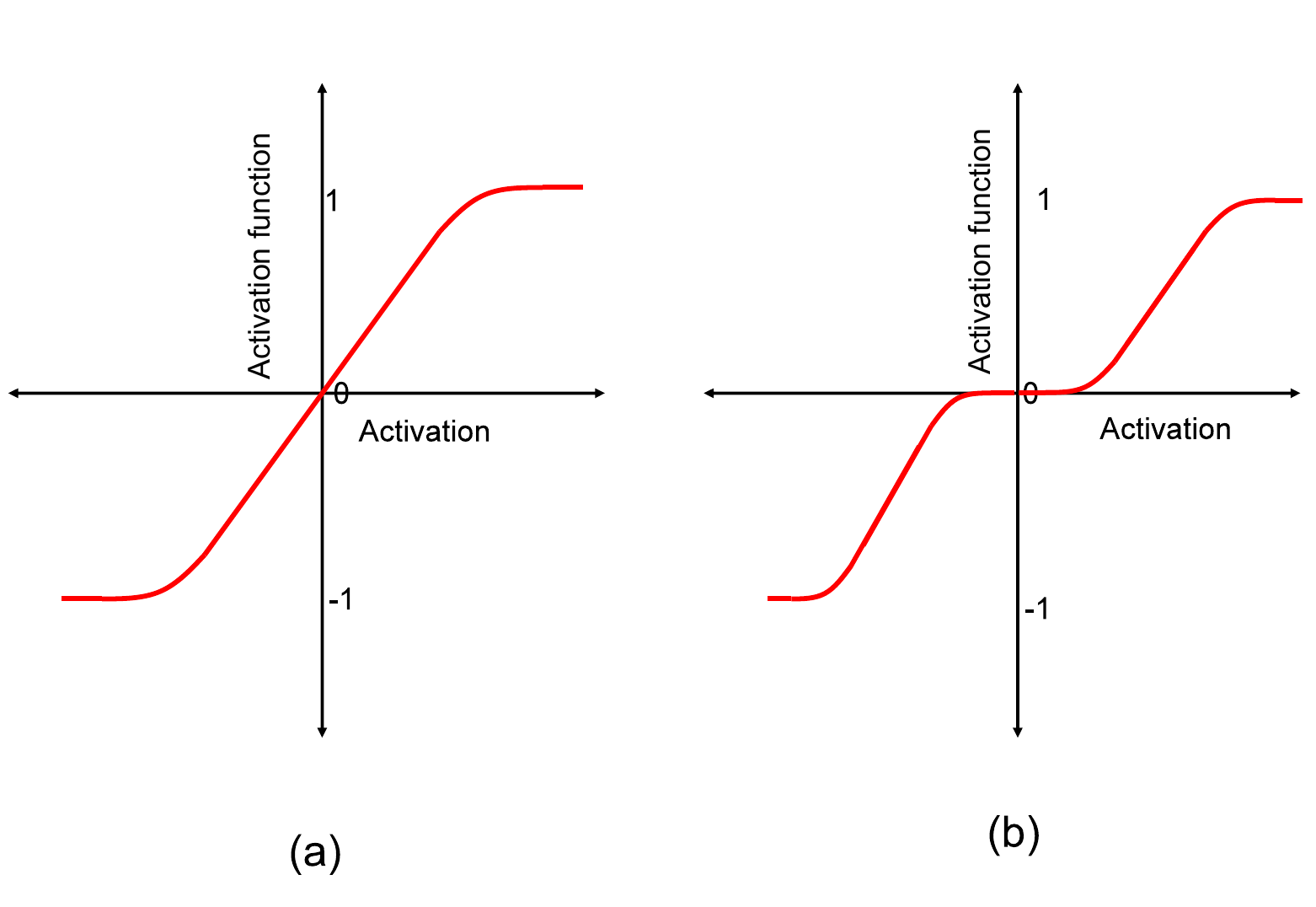}
    \caption{\small Activation function of: (a) a binary stochastic neuron, and (b) a ternary stochastic neuron. The latter looks like two staircases with a landing. The width of the ``landing'' or the plateau can be increased by increasing the magnitude of the strain.}
\label{fig:activation}
\end{figure}

Ternary stochastic neurons (TSNs) have  three output states instead of the two in BSNs. These three states can be represented by -1, 0 and +1, and the probability of being in any of these states will be manipulated by an external agent. While BSNs (with two states -1 and +1) are implemented with low-barrier nanomagnets \cite{orchi1, orchi2} and analog stochastic neurons (with any state between -1 and +1) are implemented with even lower barrier nanomagnets \cite{rahnuma}, the implementation of a TSN is more tricky. To understand the challenge, consider the activation function of a BSN which is shown in Fig. \ref{fig:activation}(a). It is a sigmoid or $tanh(x)$ function, which is very appropriate for a BSN, but not a TSN \cite{pitis}. This is because the slope near zero activation strength is highest and hence the state 0 will not be stable if such an activation function is adopted for a TSN. The neuron will tend to get pushed away from this state. Instead, we would want an activation function that looks like Fig. \ref{fig:activation}(b) which allow the intermediate state with 0 output to be stable \cite{pitis}. Ref. \cite{pitis} proposed a function to replace $tanh(x)$:
\begin{equation}
f(x) = 1.5tanh(x) + 0.5tanh(-3x), 
\end{equation}
which replicates the shape in Fig. \ref{fig:activation}(b). Needless to say that the standard low barrier nanomagnet of refs. \cite{orchi1,orchi2} cannot implement an activation function of this shape, and so we have to devise a new strategy to implement a TSN with an activation function that looks like the one in Fig. \ref{fig:activation}(b).

\begin{figure}[!h]
\centering
\includegraphics[width=4.9in,angle=270]{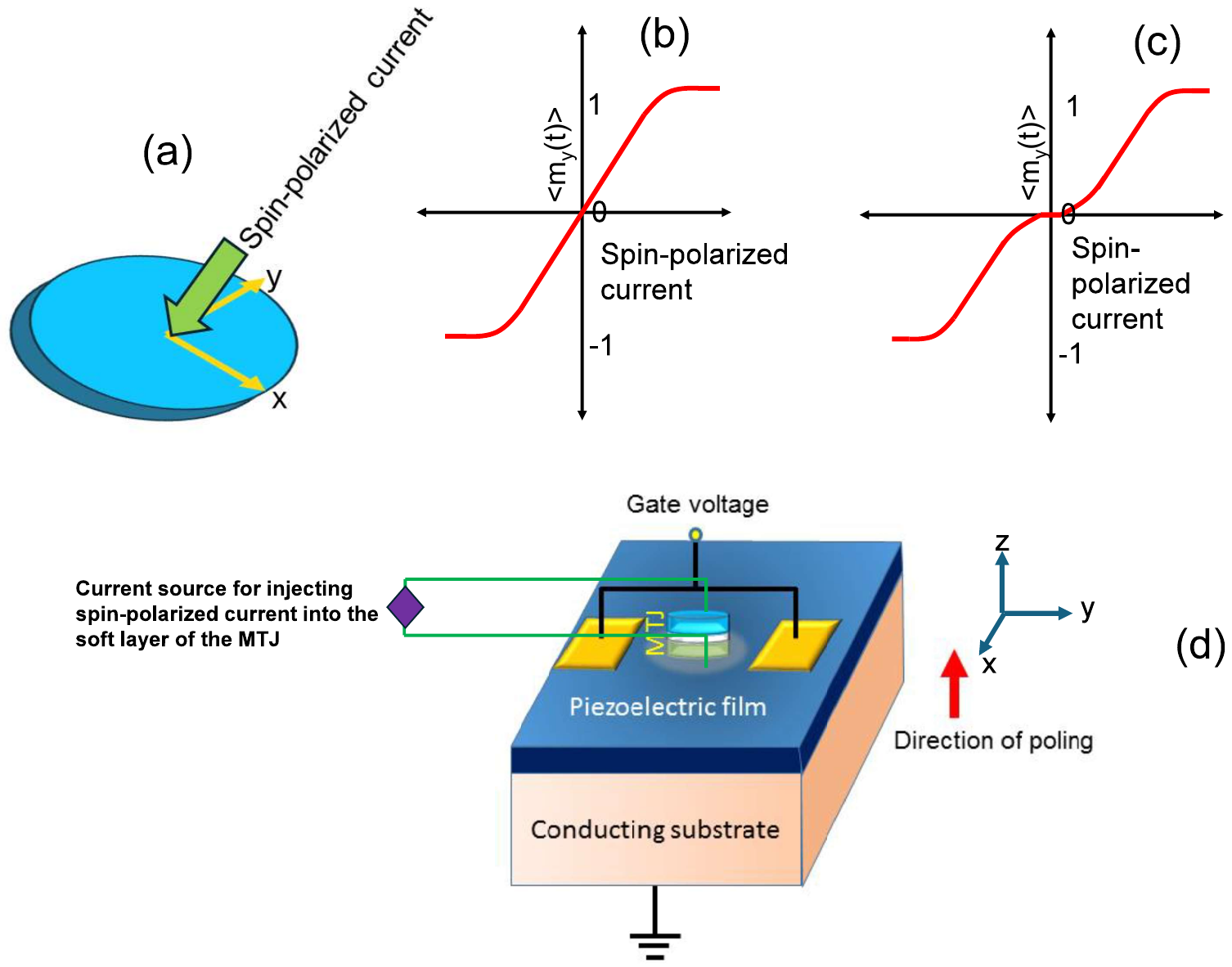}
    \caption{\small (a) A circular disk of a magnetostrictive material into which a spin-polarized current is injected perpendicular to the plane. The current is spin polarized in the $\pm$y-direction with the current's sign denoting the spin polarization and not the current polarity. (b) The y-component of the magnetization averaged over time (which is the activation function) versus the spin polarized current (which is the activation agent) when no strain is present. (c) The y-component of the magnetization averaged over time versus the spin polarized current when uniaxial strain of the correct sign is applied along the y-axis. (d) The configuration of the TSN. The magnetic tunnel junction is used to inject a spin polarized current of either polarization into the soft layer which acts as the TSN and the two shorted gate electrodes are used to apply uniaxial strain of the correct sign along the y-axis. The sign of the strain depends on the polarity of the gate voltage. If the electric field generated by the gate voltage is opposite to the direction of poling of the piezoelectric layer, then compressive (negative) strain will be generated along the axis joining the two gate electrodes (y-direction) and tensile (positive) strain will be generated in the transverse direction (x-direction). Reversing the polarity will reverse the signs of the strains. The MTJ resistance denotes the neuronal state; ``high'' is the state -1, ``intermediate'' is the state 0 and ``low'' is the state +1.}
\label{fig:disk}
\end{figure}

\section{Strained magnetostrictive nanomagnets for TSN}

Consider a magnetostrictive nanomagnet with in-plane anisotropy  made of, say, Galfenol (FeGa) that is shaped like a circular disk as shown in Fig. \ref{fig:disk}. There is no in-plane shape-anisotropy potential barrier. At room temperature, the magnetization will fluctuate randomly because of thermal perturbations and all in-plane orientations will be equally likely.  If we now inject a spin polarized current into the nanomagnet perpendicular to the plane, with the spin polarized along a chosen in-plane direction, then we will bias the probability of the magnetization to point in that direction because the spin polarized electrons in the current will transfer their momentum to the resident electrons in the nanomagnet and make the likelihood of the magnetization pointing in the direction of the spin polarization increase with increasing current strength. 

Let us assume that the current is spin polarized either along the +y-axis or -y-axis (in Fig. \ref{fig:disk}). Positive current will represent the former case and negative current the latter. The ``activation function'' is the quantity $<m_y(t)>$ averaged over time where $m_y(t)$ is the y-component of the magnetization at any instant $t$. The ``activation agent'' is the spin polarized current. If we plot the former as a function of the latter, we will obtain the $tanh(x)$-like curve of Fig. \ref{fig:activation}(a) \cite{orchi1,orchi2} which is the well-known activation curve of a BSN.

To obtain the activation function of a TSN which would look like Fig. \ref{fig:activation}(b), we will apply a unidirectional strain along the y-axis. The sign of the applied strain (tensile or compressive) will depend on the sign of the magnetostriction. The product of the strain and the magnetostriction must be {\it negative}. In that case, the uniaxial strain will tend to make the magnetization point along the x-direction for reasons we explain later. Thus, if there is no spin polarized current injection but there is uniaxial strain of the correct sign, then the magnetization will tend to point along the x-direction making $< m_y (t)>$ zero. Until there is significant spin polarized current injection which will tend to align the magnetization in the $\pm$y-direction, the magnetization will linger around the x-axis because of the strain, making $< m_y (t)>$ = 0. The strain will thus give rise to the plateau in the activation function seen in Fig.\ref{fig:activation}(b). As the current magnitude increases, the y-component of the magnetization begins to increase. This will result in the activation function shown in Fig. \ref{fig:activation}(b). This, uniaxial strain enables the activation function of a TSN.

\section{Landau-Lifshitz-Gilbert simulations to obtain the activation function versus activation strength}

We carried out Landau-Lifshitz-Gilbert (LLG) simulations of the magnetization dynamics in a circular FeGa nanomagnet of diameter 100 nm and thickness 2 nm in the presence of spin-polarized current injection and uniaxial strain. A nanomagnet of these dimensions is monodomain and hence the macrospin approximation holds. The spin polarization of the current is along the $\pm$y-direction and the uniaxial strain is also along that direction. The saturation magnetization $M_s$ = 1.32$\times$10$^6$ A/m, the magnetostriction coefficient $\lambda_s$ = 266.6 ppm and the Gilbert damping coefficient $\alpha$ = 0.017 \cite{atul}. 

The LLG equations governing the temporal evolutions of the scalar components of the magnetization normalized to the saturation magnetization $M_s$, i.e., $m_x(t)$, $m_y(t)$ and $m_z(t)$ were solved with finite difference method \cite{rahnuma1, rahnuma2} with a time step of 0.1 ps. The simulations were carried out for 1 $\mu$s and the magnetization components were sampled at every time step, yielding 10$^7$ samples to average over in order to calculate $<m_y(t)>$. 

Although the gate voltage in Fig. \ref{fig:disk}(d) will cause biaxial strain in the nanomagnet \cite{cui}, the strain components along the x- and y- axes will have opposite signs and hence reinforce each other. Therefore, we can approximate the biaxial strain as a  uniaxial strain along the y-axis with a magnitude {\it larger} than the actual magnitude.  The initial condition was that the magnetization was aligned along the -y-axis of the nanomagnet, i.e., $m_y(0)$ = -1.   

The LLG equation describing the temporal evolution of  the magnetization ${\vec m}(t)$ is:
\setlength{\mathindent}{0.3cm}
\begin{eqnarray}
    {{d {\vec m}(t)}\over{dt}} & = & - \gamma {\vec m}(t) \times \left [ {\vec H}(t) - \frac{\alpha}{\gamma} {{d {\vec m}(t)}\over{dt}}\right ] 
+ A {\vec m}(t) \times \left [{{\eta I_s(t) \mu_B}\over{q M_S \Omega}} {\hat \epsilon} \times {\vec m}(t) \right ] \nonumber \\
    && + B \left [{{\eta I_s(t) \mu_B}\over{q M_S \Omega}}{\hat \epsilon} \times {\vec m}(t) \right ].
    \end{eqnarray}
    The last term in the right-hand-side is associated with the field-like torque due to the injected spin polarized current of magnitude $I_s$ and spin polarization fraction $\eta$ ($\eta$ = 0.5)) and the second-to-last term is associated with the Slonczewski torque. Here, ${\hat \epsilon}$ is the unit vector in the direction of the spin polarization of the current. The relative magnitudes of the two torques are given by the factors $A$ and $B$. We assume that $A$ = 1 and $B$ = 0.3 \cite{kuntal}. Further \cite{fashami}, 
    \setlength{\mathindent}{2.5cm}
\begin{eqnarray} 
{\vec H}(t) & = & H_x(t) {\hat x} + H_y(t) {\hat y} +  H_z(t) {\hat z} \nonumber \\
H_x(t) & = & -M_s \frac{\pi}{4} \frac{t}{d} m_x(t) + h_x^{noise}(t) \nonumber \\
H_y(t) & = & -M_s \frac{\pi}{4} \frac{t}{d} m_y(t) + h_y^{noise}(t) + {{3}\over{\mu_0 M_s}} \lambda_s \sigma m_y(t) \nonumber \\
H_z(t) & = & -M_s \left [ 1 - 2 \frac{\pi}{4} \frac{t}{d} \right ]  m_z(t) + h_z^{noise}(t) 
\end{eqnarray}
where $t$ is the thickness of the nanomagnet, $d$ is the diameter, $\alpha$ is the Gilbert damping factor of the nanomagnet material, $\gamma$ is the gyromagnetic factor (a constant),  $h_i^{noise}(t) = \sqrt{{{2 \alpha kT}\over{\gamma \left (1 + \alpha^2 \right ) \mu_0 M_s \Omega \Delta t}}}G_{(0,1)}^i (t)$ with $G_{(0,1)}^i (t)$ ($i = x, y, z$) being three uncorrelated  Gaussians of zero mean and unit standard deviation, $\Omega$ is the nanomagnet volume, $\sigma$ is the applied stress,  and $\Delta t$ is the attempt period which is the time step of the simulation.

\begin{figure}[!ht]
    \centering
\includegraphics[width=5.3in,angle=270]{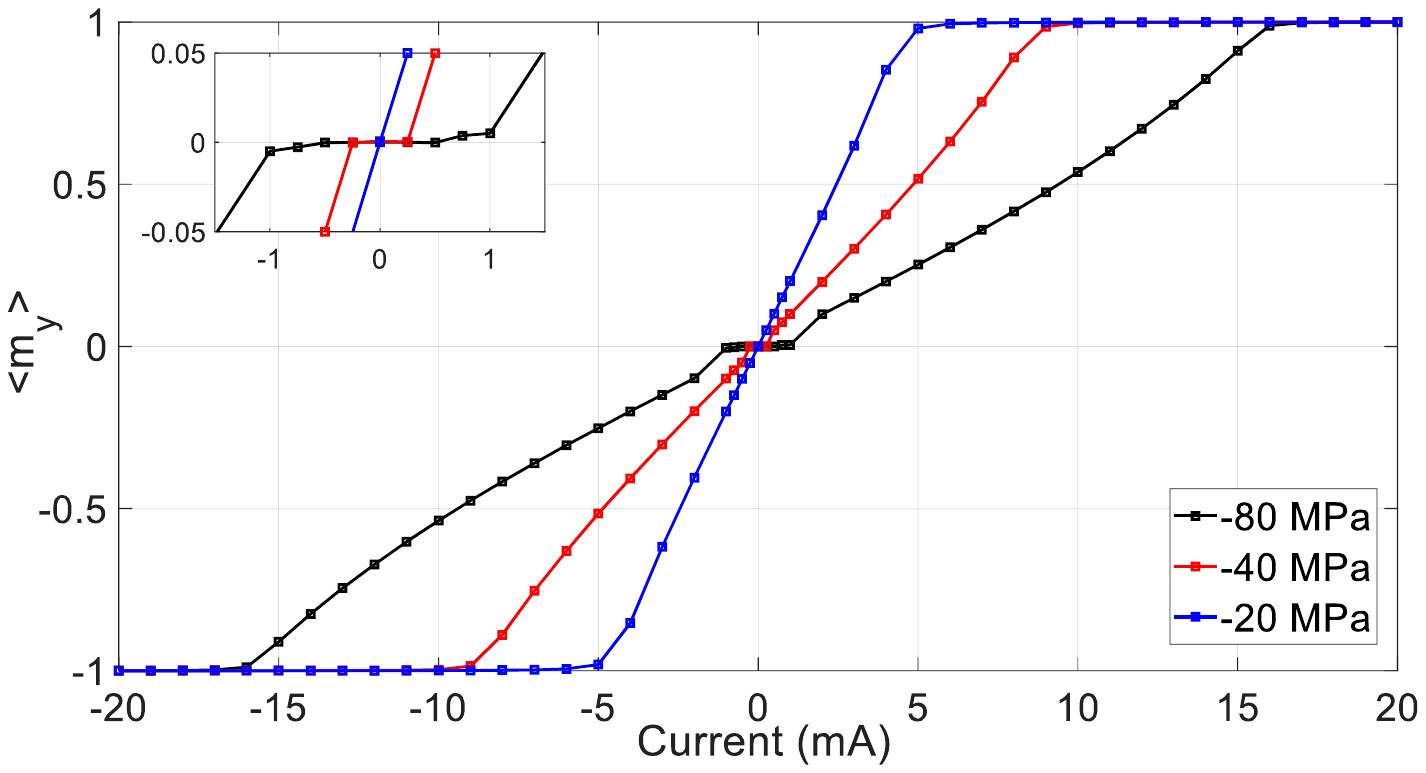}
    \vspace{-0.9in}
    \caption{\small The activation function $<m_y(t)>$ as a function of the activation strength which is the spin polarized current $I_s$ for three different compressive stress values. Positive sign of $I_s$ corresponds to current having spin polarization along the +y-direction and negative sign corresponds to the current having spin polarization along the -y-direction. Since FeGa has positive magnetostriction, a negative stress. i.e., a compressive stress, will result in the product of the stress and magnetostriction ($\lambda_s \sigma$ product) having a negative sign. Note that there is a plateau in the characteristic around zero current (shown in more detail in the inset) for stress values of 40 and 80 MPa, but not  20 MPa. Existence of this plateau is what is needed for TSN implementation.}
    \label{fig:result1}
\end{figure}

\begin{figure}[!ht]
    \centering
\includegraphics[width=\textwidth]{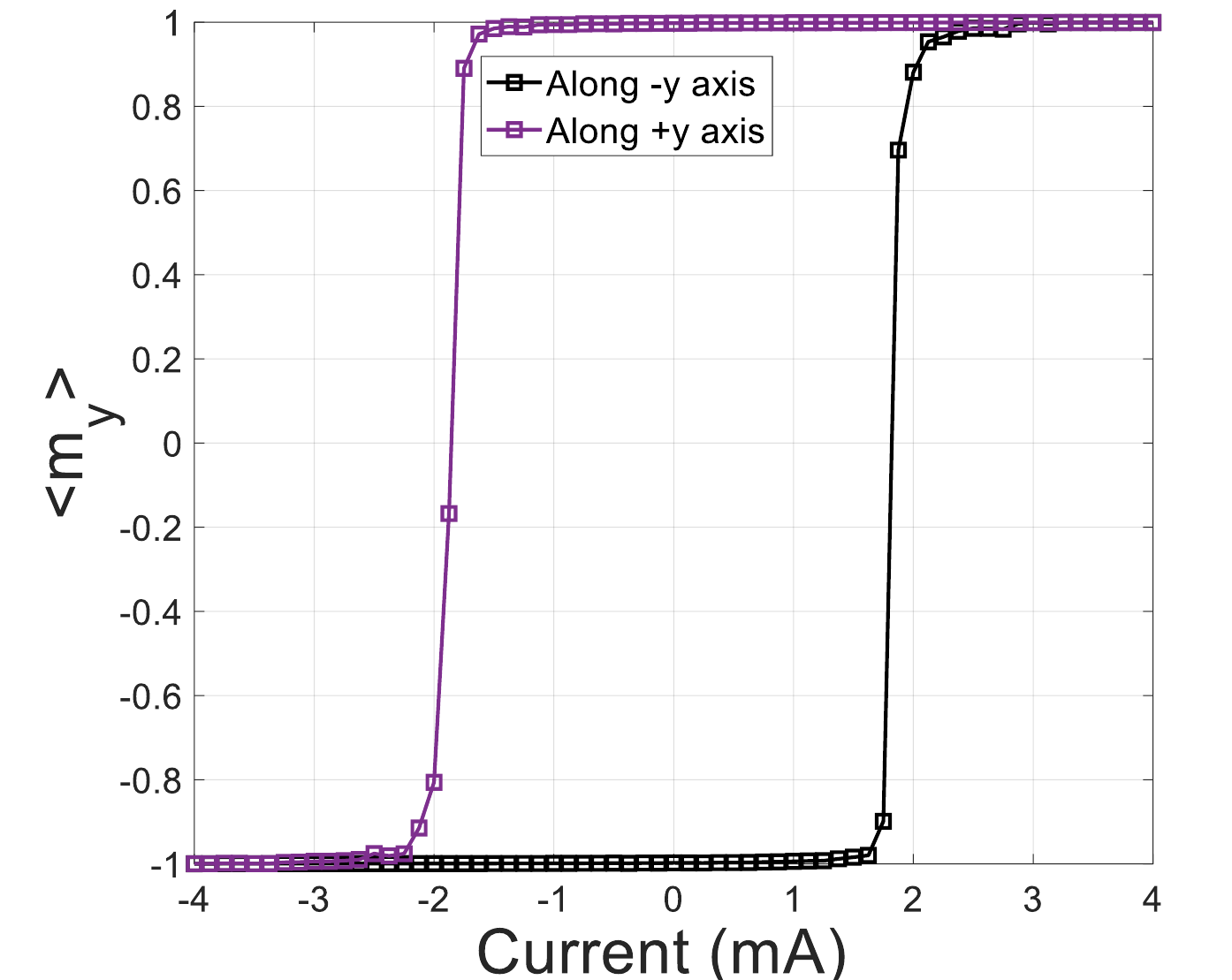}
    \caption{\small The activation function $<m_y(t)>$ as a function of the activation strength which is the spin polarized current $I_s$. Again, positive sign of $I_s$ corresponds to current having spin polarization along the +y-direction and negative sign corresponds to the current having spin polarization along the -y-direction. The results are plotted for tensile  stress of +80 MPa when the $\lambda_s \sigma$ product has a positive sign. Here, the activation function depends on the initial magnetization orientation and reason is explained in the text. We show the two cases where the initial magnetization orientation is along the +y-direction and the -y-direction. This case is obviously not useful for a TSN but is included for the sake of completeness.}
    \label{fig:result2}
\end{figure}

We pick different values of the spin polarized current $I_s$ and two different signs of the spin polarization $\eta = \pm$0.5. For each of these picks, we run the simulation for 1 $\mu$s and collect statistics at every time interval of 0.1 ps, resulting in 10$^7$ samples for every pick. We then average over the 10$^7$ values to obtain the time averaged value of $<m_y(t)>$ to obtain the activation function. We plot this quantity as a function of the activation strength, which is the spin-polarized current $I_s$. The result is shown in Fig. \ref{fig:result1} for compressive stress ($\lambda_s \sigma$ product is negative) where positive values of $I_s$ correspond to spin-polarization in the +y-direction and negative values correspond to spin-polarization in the -y direction. The result for tensile stress ($\lambda_s \sigma$ product is positive) is shown in Fig. \ref{fig:result2}. 

The maximum stress values we consider are $\pm$80 MPa. The Young's modulus of FeGa is about 75 GPa \cite{flatau}. Hence the  strain that will result from the applied stress is $\pm$80 MPa/75 GPa = $\pm$1.06$\times$10$^{-3}$ or $\pm$0.1\%, which is probably the limit of strain we can reasonably generate.

\section{The effect of uniaxial stress on the activation function}

\subsection{Positive $\lambda_s \sigma$ product or compressive stress}

The effect of stress in the absence of any spin-polarized current (or in the presence of small spin-polarized current) can be understood from the third line of Equation (3) which provides the expression for the effective magnetic field due to stress. This field is along the y-direction since uniaxial stress is applied along the y-axis. When the $\lambda_s \sigma$ product is positive, positive values of $m_y(t)$ (i.e., magnetization pointing along +y-direction) will make the effective magnetic field also point in the +y-direction [obvious from Equation (3)] which will help to keep the magnetization oriented along the +y-direction. Similarly, for negative values of $m_y(t)$, the effective magnetic field will point in the -y-direction which will again keep the magnetization oriented in the -y-direction. Thus, stress in this case will help to keep the magnetization oriented along the y-axis, either +y or -y, resulting in a non-zero value of $<m_y(t)>$ when $I_s$ $\approx$ 0.

This effect also explains why the activation function will be asymmetric along the vertical axis ($I_s$ = 0). Until there is substantial spin-polarized current injection to overcome the effect of stress, the magnetization will remain pointing along the initial orientation regardless of whether it is along the +y-direction or the -y-direction.

That begs the question what will happen if the magnetization is initially along some arbitrary direction. When the nanomagnet is stressed, the magnetization will settle into either the +y or the -y direction (with equal probability) because those locations correspond to the potential energy minima when $\lambda_s \sigma > 0$. Therefore, when the spin polarized current is injected later, the initial magnetization will always have been along either the +y-direction or the -y-direction.

\subsection{Negative $\lambda_s \sigma$ product or tensile stress}

When the   
$\lambda_s \sigma$ product is negative, positive values of $m_y(t)$ will cause the effective magnetic field due to stress to point in the -y-direction, i.e., opposite to the magnetization, and that will tend to flip the magnetization from +y to -y direction. The same effect will be observed for negative values of $m_y(t)$. Thus, stress will prevent the magnetization from settling into either orientation along the y-axis, regardless of the initial orientation, making $<m_y(t)>$ $\rightarrow$ 0 when the spin-polarized current is {\it zero or too small to overcome the effect of stress}. This is the cause for the plateau in the activation function observed in Fig. \ref{fig:result1} for compressive stress ($\lambda_s \sigma <0$), which is crucial for a TSN.   It is also obvious that larger the magnitude of stress, wider will be the plateau since a larger stress will require a larger spin-polarized current to overcome its effect. This is clearly seen in Fig. \ref{fig:result1}.

This last feature raises another important question. Clearly in an ensemble of nanomagnets acting as TSNs, there will inevitably be some stress non-uniformity across the ensemble and hence variations in the width of the plateau in the TSNs. This is not a serious drawback since as long as there is a plateau, we will have a stable state around $I_s = 0$ and the TSN functionality will be unimpaired. However, what is more serious is that the {\it shape} of the activation function also changes with stress. This shape variability has some undesirable effects such as variations in the rate of convergence of the root mean square error in training a neural network \cite{rahnuma2}. However, this effect can also accrue from variations in the size and shape of the nanomagnets \cite{rahnuma3} and is unavoidable anyway.

\begin{figure}[!h]
    \centering
\includegraphics[width=\textwidth]{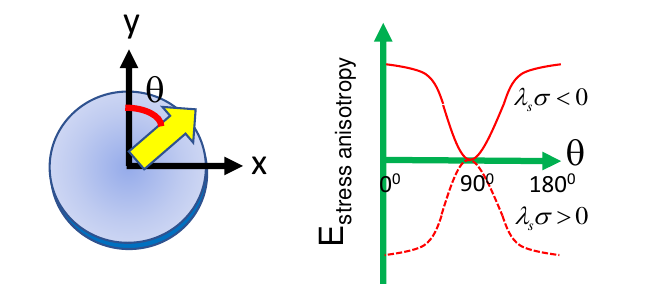}
    \caption{\small The stress anisotropy energy as a function of the magnetization orientation (shown by the yellow arrow) for both positive and negative $\lambda_s \sigma$ product. For the two stress values considered here, 50 MPa and 100 MPa, the potential hill or well will have magnitudes of 9.85 and 19.7 kT (T = 300 K).}
    \label{fig:plot}
\end{figure}

This physics may be even easier to understand  from potential energy considerations. The stress-anisotropy energy in the nanomagnet's plane is given by \cite{fashami}
\setlength{\mathindent}{3cm}
\begin{equation}
E_{stress-anisotropy} = -(3/2) \lambda_s \sigma \Omega cos^2 \theta,
\end{equation}
where $\theta$ is the angle between the magnetization orientation and stress axis (which is the y-axis in our case). If we plot this energy versus $\theta$, as in Fig. \ref{fig:plot}, we see the energy minimum occur at $\theta = 90^{\circ}$ for negative $\lambda_s \sigma$ product and at $\theta = 0^{\circ}, 180^{\circ}$ for positive 
$\lambda_s \sigma$ product.

 Hence, the negative $\lambda_s \sigma$ product will tend to align the magnetization along the x-axis ($\theta = 90^{\circ}$) making $m_y$ = 0, whereas the positive product will tend to align the magnetization along the $\pm$y-direction ($\theta$ = 0$^{\circ}$ or 180$^{\circ}$), making $m_y$ = $\pm$1. This again explains why $<m_y(t)>$ is zero when $I_s \rightarrow$ 0 for the case of negative $\lambda_s \sigma$ product and  why $<m_y(t)>$ is non-zero when $I_s \rightarrow$ 0 for the case of positive $\lambda_s \sigma$ product.

 It should be obvious to the reader that if we rotate the stress axis by 90$^{\circ}$ and align it along the x-direction, while keeping everything else the same, we will have to reverse the signs of the stresses to get the desired activation function.

\subsection{The effect of initial conditions}

For the \underline{negative $\lambda_s \sigma$ product} (compressive uniaxial stress), the initial condition does not make a difference. Regardless of what the initial magnetization orientation is, stress will make it point along the x-axis in the absence of spin-polarized current since the energy minimum is at $\theta$ = 90$^{\circ}$. Spin polarized current will have to lift it out of the energy minimum and make it point along either +y or -y-direction. Since there is no preference for either of these two directions, the plot in Fig. \ref{fig:result1} is symmetric about the vertical axis (which corresponds to $I_s$ = 0).

For the \underline{positive $\lambda_s \sigma$ product} (tensile uniaxial stress), however, the initial condition does make a difference. For example, if the magnetization is initially along the -y-direction, then in the absence of any spin polarized current, it will remain in that direction (since it is a stable orientation) and $<m_y(t)>$ will be negative when $I_s \rightarrow$ 0. Similarly, if the initial orientation is along the +y-direction, it is again at a stable orientation and hence $<m_y(t)>$ will be positive when $I_s \rightarrow$ 0. We see this trend clearly in Fig. \ref{fig:result2}. This case is not useful for TSN implementation, but we include it here for the sake of completeness. 

Note that in the case of tensile stress, the activation energy plot is {\it asymmetric} about the vertical axis. There are many asymmetric activation functions that have found use in neural networks, such as ReLU \cite{hinton}, Swish \cite{ramachandran}, Mish \cite{misra}, EANAF \cite{chai}, etc., which look similar to the activation function in Fig. \ref{fig:result2}. Application of this type of activation function for learning and inference tasks in neural networks is, however, outside the scope of this paper on TSNs and hence will be discussed elsewhere. 

\section{Threshold based ternary functions}

We also notice that  in the case of negative (compressive) stress as seen in Fig. \ref{fig:result1} the activation function can be written as
\begin{equation}
    <m_y(t)>  = \left \{
    \begin{array}{ccc}
      +1 & if & I_s > I_{s0} \\ 
      0 & if & |I_s| \le I_{s1} \\
      -1 & if & I_s < -I_{s0},
    \end{array}
\right \}
\end{equation}
 where $I_{s0}$ and $I_{s1}$ are some threshold values of the spin polarized current. This has an application in threshold based ternary functions of the nature discussed in \cite{li}. Such behavior is used to optimize the Euclidean distance between full precision and ternary weights in ternary neural networks \cite{li}. Here we find this feature manifested quite prominently in the presence of stress of the correct sign that will make the $\lambda_s \sigma$ product negative.

\section{Conclusion}

We showed how to implement the activation function of a ternary stochastic neuron (TSN) with a stressed low-barrier magnetostrictive nanomagnet injected with a spin-polarized current. To our knowledge, this is the first and only nanomagnetic implementation of a TSN. There are no serious challenges to applying stress to alter the potential energy landscape of a magnetostrictive nanomagnet in the fashion outlined here with extremely small energy dissipation since it has been demonstrated repeatedly over the past decade \cite{book,APR,book2}.

The BSN and the TSN will have the same area since they are both implemented with the same nanomagnet (the difference is that one is stressed and the other is not), but the TSN has more data processing power embedded in the same area, which is a boon in the age of data deluge. Comparatively speaking, TSNs will also reduce energy cost, number of interconnects and increase information content per unit area. In terms of area comparisons, the addition of the gate peripherals to apply stress will increase area, but the same gate pads can apply stress {\it on all the nanomagnets} simultaneously and hence the increase, when amortized over numerous nanomagnets, is miniscule. This method of implementing a TSN, which is simple and effective, can benefit edge intelligence and embedded applications.

We point out that that the contribution here is not just with respect to the activation function. We have {\it enabled} nanomagnetic tristate stochastic neural networks and thereby improved Ising machines, Boltzmann machines, Bayesian inference engines and anything else that can employ stochastic neurons by enabling the first thing that’s required, namely a 3-state activation function. This is a key enabler of $n$-ary stochastic neural networks and has far-reaching implications. 

Finally, we address an important point. To implement the TSN, we have to keep the strain on all the time. Since the strain is generated with a voltage on the piezoelectric, the voltage must be kept on all the time (the construct is ``volatile'') and it might appear that this will cause unacceptable standby power dissipation, but that is not correct. The piezoelectric acts as a capacitor which is fully charged at any given voltage. There is no more charging or discharging current once steady state is reached, and hence there is no standby power dissipation. The charges will of course ultimately leak out with time and some refresh cycles will be needed, but this is not `standby power dissipation' and is not exorbitant since it is the standard practice in dynamic random access memory.  Furthermore, because the strain is {\it constant} and not variable, it can also be generated by depositing the nanomagnet on an appropriate lattice-mismatched substrate.  The thickness of the nanomagnet is small enough that the layer will be pseudomorphic, i.e., the strain will not be relaxed through dislocation and defects. This can eliminate the need for the voltage altogether. However, the voltage-based approach is reconfigurable while the latter is not.

\section*{Acknowledgments}
This work was supported by the US Air Force Office of Scientific Research under grant FA865123CA023 and the Virginia Innovation Partnership Corporation under grant CCF23-0114-HE.

\end{document}